\documentclass[fleqn,usenatbib]{mnras}

\usepackage{newtxtext,newtxmath}

\usepackage[T1]{fontenc}
\usepackage{ae,aecompl}

\usepackage{graphicx}	
\usepackage{amsmath}	
\usepackage{amssymb}	

\newcommand{\degrees}{$^{\rm{o}}$}
\newcommand{\ms}{\mbox{m\,s$^{-1}$}}
\newcommand{\kms}{\mbox{km\,s$^{-1}$}}
\newcommand{\Msun}{\mbox{$M_{\odot}$}}

\newcommand{\Mjup}{\mbox{$M_{\rm Jup}$}}

\newcommand{\gtsimeq}{\raisebox{-0.6ex}{$\,\stackrel
        {\raisebox{-.2ex}{$\textstyle >$}}{\sim}\,$}}

\title[Truly eccentric I]{Truly eccentric. I. Revisiting eight single-eccentric planetary systems}

\author[R.A. Wittenmyer et al.]{
Robert A. Wittenmyer,$^{1}$\thanks{E-mail: rob.w@usq.edu.au}
Jake T. Clark$^{1}$
Jinglin Zhao,$^{2}$
Jonathan Horner$^{1}$
\newauthor Songhu Wang$^{3,4}$, and Daniel Johns$^{5}$
\\
$^{1}$University of Southern Queensland, Centre for Astrophysics, USQ Toowoomba, QLD 4350 Australia\\
$^{2}$UNSW School of Physics, Sydney, NSW 2052 Australia\\
$^{3}$Department of Astronomy, Yale University, 52 Hillhouse Avenue, New Haven, CT 06511, USA \\
$^{4}$51 Pegasi Fellow \\
$^{5}$Department of Physical Sciences, Kutztown University, Kutztown, PA 19530, USA
}

\date{Accepted XXX. Received YYY; in original form ZZZ}

\pubyear{2018}

\hypersetup{draft} 
\begin{document}
\label{firstpage}
\pagerange{\pageref{firstpage}--\pageref{lastpage}}
\maketitle

\begin{abstract}
We examine eight known single-eccentric planetary systems in light of recently released large data archives and new analysis techniques.  For four of these systems (HD\,7449, HD\,65216, HD\,89744, HD\,92788) we find evidence for additional long-period companions.  HD\,65216c is a Jupiter analog, with a period of 14.7\,yr, $e=0.18$, and m sin $i$ of 2\Mjup, whilst the remaining candidate companions move on as-yet-incomplete orbits. Our results highlight the importance of revisiting the analysis of known exoplanetary systems when new data become available, particularly given the possibility that poorly-sampled data might previously have led to the detection of a 'false-positive' single eccentric planet, when the system in question actually contains two (or more) planets on near-circular orbits.

\end{abstract}




\section{Introduction}

In the past thirty years, the discovery of planets orbiting other stars has moved from being an unfulfilled dream to an apparently routine process. The first planets discovered orbiting Sun-like stars were discovered using the radial velocity (RV) technique \citep[e.g.][]{campbell88,lathamsworld,51peg,early1}, which dominated the discovery of new planets for the first two decades of the Exoplanet Era, and still plays a vital role in our ongoing search for alien worlds \citep[e.g.][]{bonfils13,fischer16, butler17}.

The RV technique enables astronomers to probe a unique region of the exoplanetary discovery phase space. Whilst it is most sensitive to massive planets moving on short-period orbits, if observations of a given star are continued over a long enough period, the technique can be used to find Jupiter- and Saturn-analogues: massive planets moving on orbits with periods measured in decades, rather than days \citep[e.g.][]{z13,2jupiters,santos16,endl16}. To do this requires observations covering most, if not all, of a complete orbital period of the planet in question.

For this reason, legacy RV surveys such as the Anglo-Australian Planet Search \citep[e.g.][]{carter03,jones10,tinney11,30177} offer a unique window to the true diversity of exoplanetary systems, and currently represent a significant resource for understanding the degree to which the Solar system is unique \citep[e.g.][]{etaearth,fressin13,jupiters,bryan18}.

Although such surveys deliver long-period coverage of their target stars, the data they yield is typically only sparsely sampled, with just a few observational epochs per year, particularly in the latter stages of a survey's lifetime \citep{witt13pasp}. Whilst such observations are more than adequate to identify the presence of planets moving on long-period orbits, the analysis of such data is fraught with unusual challenges related to the sparsely sampled nature of the data.

Of particular interest is the question of the multiplicity in planetary systems discovered using sparse RV data.  With limited data, it can be possible to miss the presence of additional planets, by focusing too heavily on a single, dominant signal. At the same time, it is possible to find 'planets' that do not actually exist, by identifying periodicities in the data that either vanish with the acquisition of more observations, or turn out to be the result of  other astrophysical phenomena \citep[such as stellar activity, e.g.][]{MurderDeathKill, nokapteyn, byebyeBb, hatzes18, diaz18}.


Science is the pursuit of truth.  In that spirit, it remains critically important to revise our understanding of the architectures of planetary systems as new data become available.  There is a vast body of literature concerning the analysis of RV data for planetary signals in the midst of the confounding effects of sparse sampling and stellar noise.  Of particular relevance is work that has re-analysed data on known planetary systems to confirm, clarify, or refute the published planetary parameters \citep[e.g.][]{df10, jenkins14, kurster15, trifonov17}.  One well-known problem is the degeneracy between a single eccentric planet and two circular planets in 2:1 period commensurability \citep[e.g.][]{ang10, songhu, boisvert18}.  In a companion paper (Wittenmyer et al., MNRAS submitted), we examine this idea is some detail, using simulated data to study the variety of ``false-positive'' single planets moving on eccentric orbits that can be mistakenly identified as a result of poorly sampled observations of such a deceptive couplet. We find that the such ``false-positives'' typically occupy a ``danger-zone'', with orbital eccentricities between $\sim 0.21$ and $\sim 0.46$ accounting for 95\% of all cases of mistaken identity for a pair of planets in, or close to, mutual 2:1 mean-motion resonance.

Following this logic,
\citet{songhu} investigated 82 known single-eccentric ($e>0.3$) planetary systems, and explored the possibility that the observed RV data could be better fit with two planets in nearly-circular orbits.  They found a subset of nine systems for which a two-planet solution reasonably fit the data and satisfied dynamical stability tests.  In this work, we revisit those systems in light of newly available data and new analysis techniques.

This paper is organised as follows: Section~\ref{Data} describes the RV data used here and our analysis approach.  Section~\ref{sec:results} gives the results of the updated orbital solutions and discusses the new planets in further detail.  Finally, we give our conclusions in Section~\ref{Conclusions}.

\section{Data and Analysis Techniques}
\label{Data}

In this section, we describe the provenance of the various data sets used in our analysis.  We also detail the techniques used to identify and fit Keplerian orbits to the RV data, as well as techniques to test the veracity of the derived signals.

\subsection{Radial Velocity Data}

The RV data were obtained largely from the public releases of the Lick and Keck planet search programs \citep{fischer14, butler17}, and the ESO archive of publicly-available HARPS spectra.  The Keck data archive was recently corrected for several small but noticeable systematic effects by \citet{talor18}, and here we use their corrected velocities for HD\,3651, HD\,52265, and HD\,92788.  Table~\ref{tab:rvdata} summarises the origin and properties of the data used herein.  No fresh observational data were available for the ninth system noted in \citet{songhu}, GJ\,649, and hence that system was not investigated in this work.  Table~\ref{tab:stars} gives a brief summary of the stellar properties for the eight stars considered here.


For HD\,117618, we included previously unpublished data from the 3.9m Anglo-Australian Telescope; the complete set of 79 AAT velocities is now given in Table~\ref{tab:117618vels}.  The HARPS fibre feed was upgraded on JD 2457161 \citep{locurto15}, and hence we treat all data after that point as coming from a separate instrument with its own velocity offset as a free parameter in the fitting process.  For all data sets, where there were multiple observations in a single night, we binned them together using the weighted mean value of the velocities in each night.  We adopted the quadrature sum of the rms about the mean and the mean internal uncertainty as the error bar of each binned point.

\begin{table}
	\centering
	\caption{Summary of Stellar Parameters}
	\label{tab:stars}
	\begin{tabular}{lllll} 
		\hline
		Star & T$_{eff}$ & log $g$ & Mass  & Reference \\
		     &  K        & cgs     & \Msun  & \\
		\hline
        HD 3651     & 5221 & 4.45 & 0.799 & \citet{vf05} \\
        HD 7449     & 6024 & 4.51 & 1.053 & \citet{sousa08} \\
        HD 52265    & 6136 & 4.36 & 1.204 & \citet{sousa08} \\
        HD 65216    & 5612 & 4.44 & 0.874 & \citet{sousa08} \\
        HD 85390    & 5186 & 4.41 & 0.758 & \citet{sousa08} \\
        HD 89744    & 6291 & 4.07 & 1.860 & \citet{vf05} \\
        HD 92788    & 5744 & 4.39 & 1.032 & \citet{sousa08} \\
        HD 117618   & 5990 & 4.41 & 1.077 & \citet{sousa08} \\
		
		
\hline
	\end{tabular}
\end{table}

\subsection{Orbit Fitting}



To fit the RV data, we obtained Bayesian posterior distributions of each planetary system's orbital parameters using the Markov Chain Monte Carlo (MCMC) code Exoplanet Mcmc Parallel tEmpering Radial velOcity fitteR\footnote{\href{https://github.com/ReddTea/astroEMPEROR}{https://github.com/ReddTea/astroEMPEROR}} ({\sc astroemperor}, Jenkins \& Pena, in prep.). As described in \citep{76920}, {\sc astroemperor} utilises thermodynamic integration methods \citep{g05} backed by an affine invariant MCMC engine, deployed by the {\sc Python} {\tt emcee} package \citep{fm13}. Using an affine invariant algorithm such as {\tt emcee} allows the MCMC analysis to perform equally well under all linear transformations consequently being insensitive to covariances among the orbital-fitting parameters \citep{fm13}. 

A model selection is performed automatically by {\sc emperor}, whereby an arbitrary posterior comparison and threshold Bayes factor of 5 and 150 respectively is required for a $k+1$ planet model to be better favoured than the previous $k$ planet model. The {\sc astroemperor} code also automatically determines which of the orbital parameters, such as period and amplitude, are statistically significantly different from zero, with the Bayesian information criterion (BIC), Akaike information criterion (AIC) and maximum a posteriori probability (MAP) estimate values calculated for each planetary signal. Flat priors are applied to all parameters except for the eccentricity and jitter priors that are folded Gaussian and Jeffries, respectively.
The Bayesian phase-space for {\sc astroemperor} can be bounded by the following parameters; $\sqrt[]{e}\cos\omega$ , $\sqrt[]{e}\sin\omega$ , $\sqrt[]{K}\cos\phi$ , $\sqrt[]{K}\sin\phi$ and $P$ where $e$, $\omega$, $K$, $\phi$, and $P$ are the orbital eccentricity, argument of periastron, semi-amplitude, mean anomaly phase and the orbital period. The mean anomaly phase is somewhat unique to {\sc astroEMPEROR} and is related to the mean anomaly $M$, $P$, and epoch time $t$ by: 

\begin{equation*}
	\centering
	\phi = M - \frac{2 \pi t}{P}
\end{equation*}

and is related by time at periastron $T_0$ by:

\begin{equation*}
	\centering
	\phi = - \frac{2\pi T_0}{P} 
\end{equation*}

If boundaries are not defined, {\sc astroemperor} will then search that parameter-space in an unbounded manner. {\sc astroemperor} was ran with the majority of parameters being unbound except for the orbital period of each planetary signal $P_i$. A candidate planet's orbital period was bounded between the \textit{Systemic}'s $P_i - \sigma_{P_i}$ and $P_i + \sigma_{P_i}$ values. A caveat to the single boundary was HD\,7449c, which is explained further in section \ref{sec:HD7449}. For our analysis, the 'burn-in' Markov-chains were 7 million iterations long (7 temperatures, 200 walkers and 5,000 steps) with another 14 million chains exploring the parameter space thereafter (10,000 steps instead of 5,000).

We checked the consistency of our fits by next using the \textit{Systemic Console} version 2.2000 \citep{mes09}, fitting Keplerian signals using a traditional ``pre-whitening'' approach, whereby signals were identified and removed as successive peaks in the Generalised Lomb-Scargle periodogram \citep{zk09}.  The results from \textit{Systemic} were entirely consistent with those from {\sc astroemperor}. 

\begin{table}
	\centering
	\caption{Summary of RV Data}
	\label{tab:rvdata}
	\begin{tabular}{llll} 
		\hline
		Star & $N_{RV}$ & Source & Reference \\
		\hline
		HD 3651 & 4 & 2.7m/CS23 & \citet{witt09} \\
        HD 3651 & 35 & HET/HRS & \citet{witt09} \\
		HD 3651 & 155 & Lick & \citet{fischer14} \\
		HD 3651 & 89 & Keck/HIRES & \citet{butler17} \\
		HD 7449 & 117 & HARPS & \citet{dum11} \\
		HD 52265 & 91 & CORALIE & \citet{naef01} \\
		HD 52265 & 66 & Keck/HIRES & \citet{butler17} \\
		HD 52265 & 4 & HARPS & ESO Archive \\
		HD 65216 & 52 & CORALIE & \citet{mayor04} \\
		HD 65216 & 24 & HARPS & ESO Archive \\
		HD 85390 & 114 & HARPS & \citet{mordasini11} \\
		HD 89744 & 9 & 2.7m/CS23 & \citet{witt09} \\
		HD 89744 & 33 & HET/HRS & \citet{witt09} \\
		HD 89744 & 117 & Lick & \citet{fischer14} \\
		HD 89744 & 74 & AFOE & \citet{korzennik00} \\
		HD 92788 & 53 & CORALIE & \citet{butler17} \\
		HD 92788 & 40 & Lick & \citet{fischer14} \\
		HD 92788 & 41 & Keck/HIRES & \citet{butler17} \\
		HD 92788 & 12 & HARPS & ESO Archive \\
		HD 117618 & 23 & HARPS & ESO Archive \\
		HD 117618 & 79 & AAT/UCLES & This work \\
\hline
	\end{tabular}
\end{table}

\begin{table}
	\centering
	\caption{AAT Radial Velocities for HD 117618}
	\label{tab:117618vels}
	\begin{tabular}{lll} 
		\hline
		BJD-2400000 & Velocity (m/s) & Uncertainty (m/s) \\
		\hline
50831.18597  &     -11.6  &    2.4  \\
50917.10104  &       8.7  &    3.6  \\
50970.94927  &      15.8  &    2.8  \\
51212.20608  &     -13.5  &    3.1  \\
51236.22669  &       1.5  &    6.7  \\
51274.24420  &      -0.7  &    3.7  \\
51383.93108  &       1.4  &    2.5  \\
51386.85838  &       1.8  &    2.5  \\
51631.25935  &     -29.6  &    2.4  \\
51682.97674  &     -16.3  &    2.8  \\
51718.03450  &       3.0  &    2.9  \\
51920.26309  &       4.8  &    3.4  \\
51984.10352  &     -18.5  &    4.2  \\
52092.96337  &     -16.5  &    2.5  \\
52129.00532  &       3.7  &    4.2  \\
52387.04015  &       7.0  &    2.0  \\
52388.07932  &      10.7  &    2.2  \\
52422.00889  &      -1.9  &    2.0  \\
52452.97666  &     -12.2  &    1.9  \\
52455.92575  &      -6.4  &    2.0  \\
52509.87274  &     -15.9  &    2.0  \\
52510.87230  &      -8.2  &    2.1  \\
52710.17772  &      -9.5  &    1.8  \\
52710.96776  &      -5.6  &    2.0  \\
52712.07593  &     -12.1  &    1.8  \\
52745.14357  &       9.3  &    2.3  \\
52750.10322  &      12.2  &    1.9  \\
52752.08884  &      12.3  &    2.0  \\
52784.00063  &       2.6  &    3.2  \\
52785.06446  &      -3.9  &    1.7  \\
52785.98809  &      -8.5  &    1.8  \\
52857.88022  &      10.9  &    1.7  \\
53006.24265  &       8.4  &    1.8  \\
53007.24120  &       1.2  &    2.4  \\
53008.23853  &       6.9  &    1.6  \\
53041.23361  &       1.6  &    2.7  \\
53042.22943  &      -7.7  &    1.9  \\
53044.16694  &     -12.6  &    2.2  \\
53045.27837  &     -15.6  &    2.1  \\
53046.16674  &     -13.8  &    1.8  \\
53047.20196  &     -16.7  &    1.8  \\
53051.19474  &      -8.2  &    2.0  \\
53213.99361  &       8.7  &    1.5  \\
53214.89440  &       6.8  &    1.7  \\
53215.89167  &       6.9  &    1.9  \\
53216.92638  &       8.2  &    1.8  \\
53242.90313  &       2.6  &    1.8  \\
53244.94739  &       2.4  &    2.2  \\
53245.88122  &       2.1  &    1.8  \\
53399.20912  &       9.8  &    1.5  \\
53405.21475  &      -7.0  &    1.5  \\
53483.04532  &     -13.7  &    2.6  \\
53485.09001  &     -10.6  &    1.9  \\
53507.02859  &      -3.2  &    1.9  \\
53521.98722  &      12.1  &    1.8  \\
53568.94943  &     -13.9  &    1.7  \\
53576.90318  &       2.2  &    1.6  \\
53943.89985  &      -6.5  &    1.3  \\
54144.17403  &       7.5  &    1.8  \\
54224.16388  &       0.2  &    1.8  \\
54254.02744  &      -2.7  &    1.6  \\
54545.13699  &     -26.9  &    1.6  \\
\hline
	\end{tabular}
\end{table}

\begin{table}
	\centering
	\contcaption{AAT Radial Velocities for HD 117618}
	\label{tab:117618velsagain}
	\begin{tabular}{lll} 
		\hline
		BJD-2400000 & Velocity (m/s) & Uncertainty (m/s) \\
		\hline
54897.21935  &       8.9  &    1.9  \\
54904.21631  &     -20.0  &    3.0  \\
55313.10475  &      -0.3  &    1.6  \\
55376.93190  &       0.1  &    1.7  \\
55402.89522  &      -8.6  &    1.9  \\
55665.16559  &      21.8  &    1.7  \\
55964.26158  &      -3.9  &    1.6  \\
56049.09986  &      15.3  &    2.0  \\
56139.90461  &      -6.2  &    1.9  \\
56379.15229  &     -22.6  &    2.5  \\
56465.95754  &      17.9  &    2.5  \\
56712.25065  &     -11.4  &    2.2  \\
56749.13034  &      11.1  &    1.9  \\
56767.03333  &      10.8  &    2.8  \\
56794.95676  &       6.9  &    1.9  \\
57054.25204  &       5.6  &    1.7  \\
57095.19944  &      -7.0  &    3.1  \\
\hline
	\end{tabular}
\end{table}

\section{Results}\label{sec:results}

\begin{figure*}
	\includegraphics[width=\columnwidth]{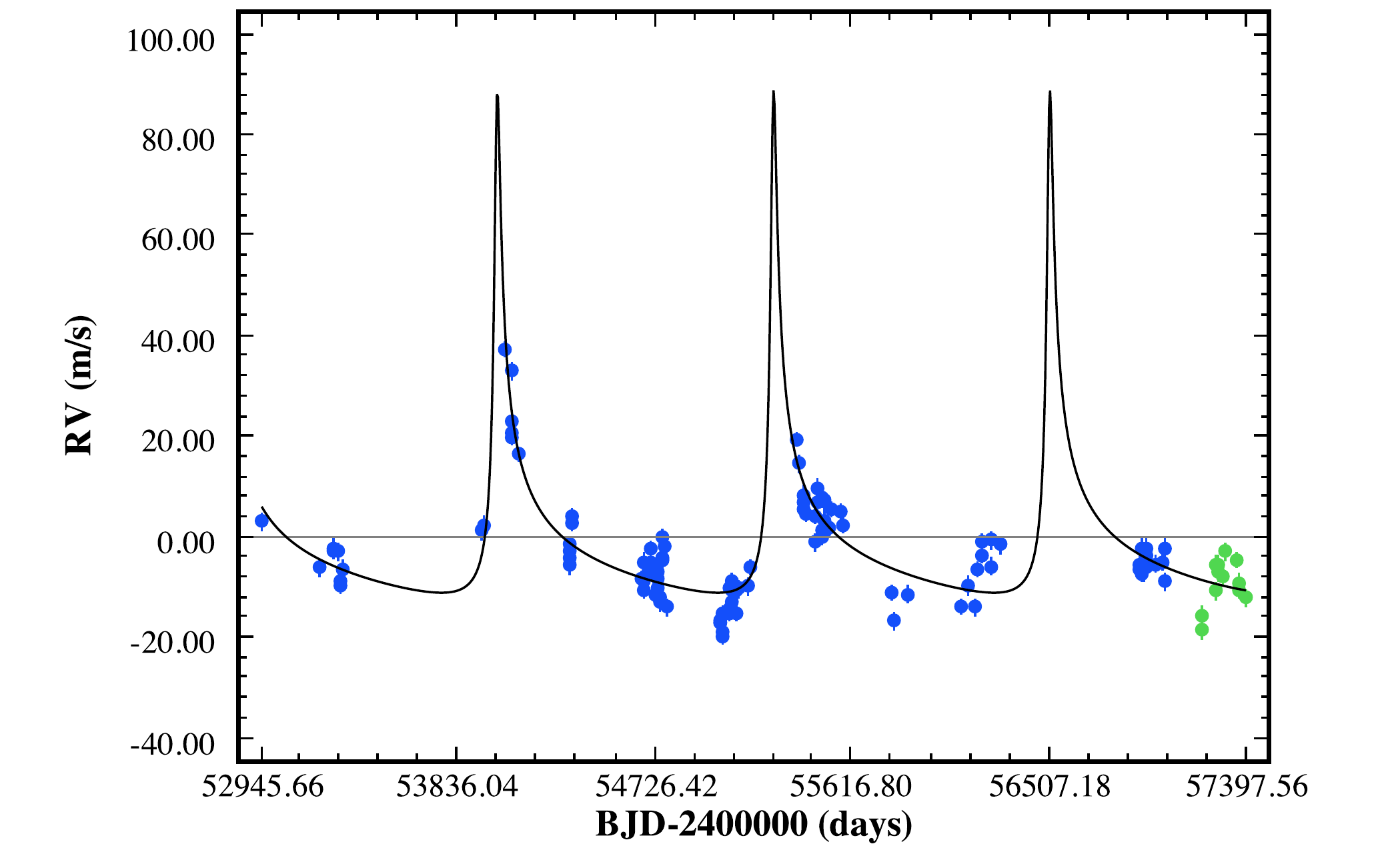}
    \includegraphics[width=\columnwidth]{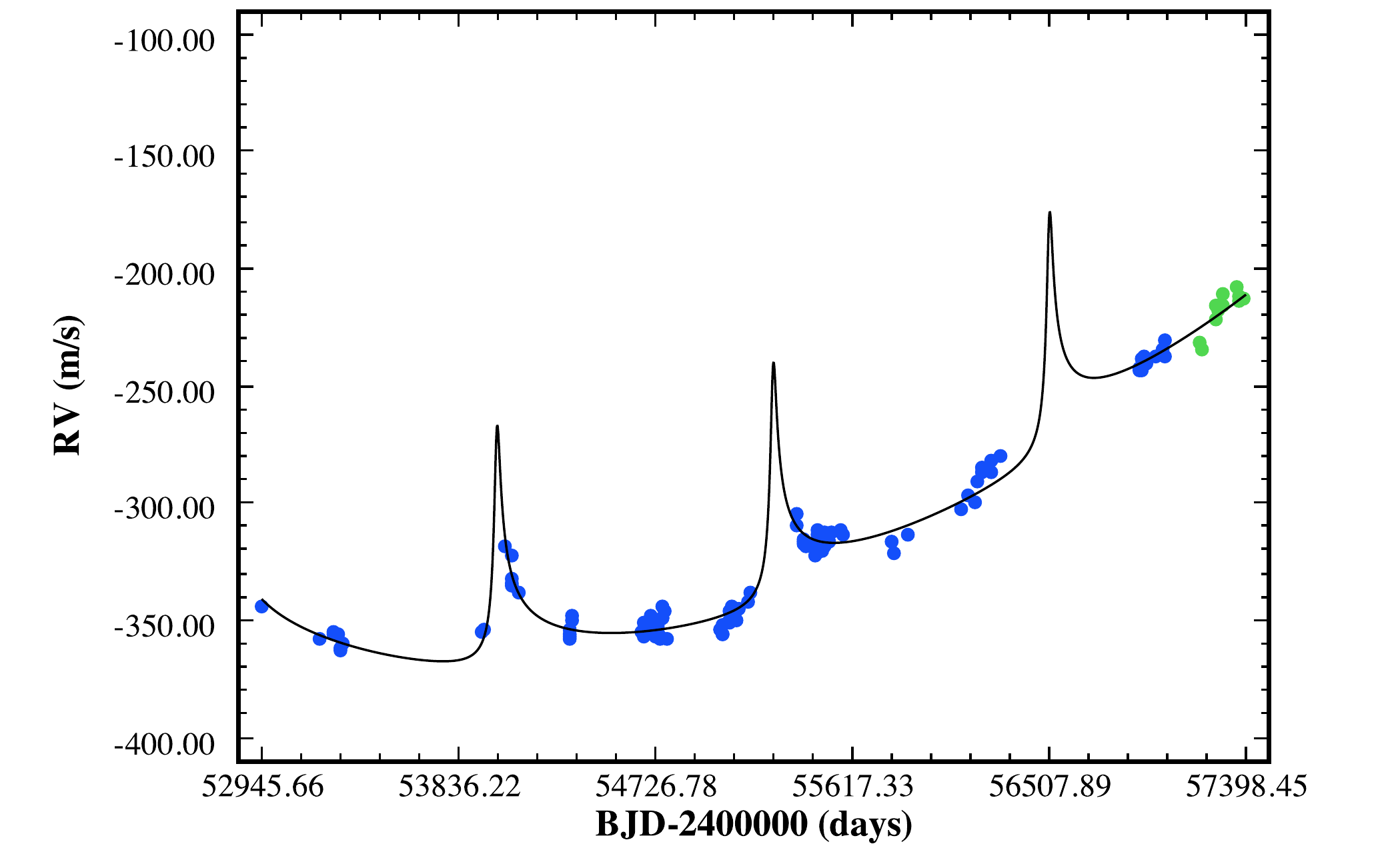}
    \caption{Left: Data and model fit for HD\,7449b, with the long-period signal removed.  Right: Data and model fit for the two bodies in the HD\,7449 system.  The outer body has $P\gtsimeq$42 yr and its eccentricity is fixed at zero. Green data are those taken after the HARPS fibre upgrade in 2015. }
    \label{fig:7449}
\end{figure*}

Our re-analysis of the eight systems considered here revealed three unconstrained long-period companions, as well as a new Jupiter analog, HD\,65216c, which is a giant planet moving on a Jupiter-like orbit at 5.75$\pm$0.10 au.  We also identify an activity-induced signal in HD\,85390, which mimics a Saturn analog in the RV curve ($P\sim$18 yr, $K\sim$3\,\ms). 
Table~\ref{tab:Bayesfit} gives our best fit solutions from {\sc astroemperor}, and Table~\ref{tab:Bayesjit} shows the fitted offsets and jitter values for each dataset.  Data and model fits are shown in Figures~\ref{fig:7449}-\ref{fig:92788} for those systems in which we find evidence for new companions, and Table~\ref{tab:bicmeasuring2} gives the BIC, AIC, and MAP values for the multi-planet {\sc astroemperor} fits.


\subsection{HD 3651}

While this system can be fit with two circular planets in a 2:1 configuration, doing so does not significantly improve the rms or goodness of fit, and does not justify the additional complexity.  We obtained a total rms of 6.54\,\ms\ about the one-planet fit using four data sets.  There is no evidence for any further incipient signals after removing the single eccentric planet orbit. 

\subsection{HD 7449}\label{sec:HD7449}

We find a long-period trend with significant curvature, to which we fit a circular Keplerian orbit as the period is too long to meaningfully suggest any eccentricity. HD\,7449c's eccentricity was bounded by {\sc astroemperor} having to explore  $\sqrt[]{e}\cos\omega$ and  $\sqrt[]{e}\sin\omega$ between 0 and $10^{-6}$. One of {\sc astroemperor}'s shortfalls is the eccentricity-periastron boundary entanglement, having to bound both $\sqrt[]{e}\cos\omega$ and $\sqrt[]{e}\sin\omega$. Due to this entanglement, {\sc astroemperor} gave a result for $\omega$ which is irrelevant for a circular orbit and has been disregarded in Table~\ref{tab:Bayesfit}. A thorough discussion of the nature of the outer companion is given in \citet{dum11}, with the added leverage of seven years of CORALIE data which unfortunately remain unpublished.  Our results support the presence of a brown dwarf with minimum mass of 31\,\Mjup\ in a poorly constrained orbit of at least $\sim$42 years.  The high-eccentricity planet reported by \citet{dum11} is then recovered, with a total rms of 4.21\,\ms\ around the two-planet solution.  Attempts to fit two circular planets in its stead failed to give convincing results; while \citet{songhu} suggested a second planet at $P\sim$615 days, the current data do not support that configuration.  Examination of 8.8 years (259 epochs) of All-Sky Automated Survey (ASAS) photometry \citep{asas1} shows no periodicities of
significance.  The ASAS $V$ band photometry has a mean value of 7.493$\pm$0.014 mag.  We note that the orbit of HD\,7449b has a gap in phase coverage near the RV peak, and that its shape is quite similar to that of the recently-discovered highly eccentric planet HD\,76920b \citep{76920}.  In that work, as in this one, the authors noted that the system's best-fit $e$ and $\omega$ could not be reproduced by the double-circular degeneracy \citep[e.g.][]{boisvert18}.  For HD\,7449b, we encourage further observations timed to capture in detail the next RV maximum, which will occur from 2020 May-July.

\subsection{HD 52265}

In addition to the CORALIE data of \citet{naef01}, we now include 66 epochs from Keck \citep{butler17} and 4 new epochs from HARPS (binned from 15 individual velocities).  As for HD\,3651, this system can be fit with two circular planets in the 2:1 configuration, but the single eccentric solution remains preferred, with a total rms of 10.1\,\ms.  Our fit is consistent with the discovery work \citep{naef01}, except that we obtain a slightly smaller eccentricity ($e=0.26\pm$0.02 here compared to $e=0.35\pm$0.03).  There is no evidence for any residual signals. 

\subsection{HD 65216}

\begin{figure*}
	\includegraphics[width=0.6\columnwidth]{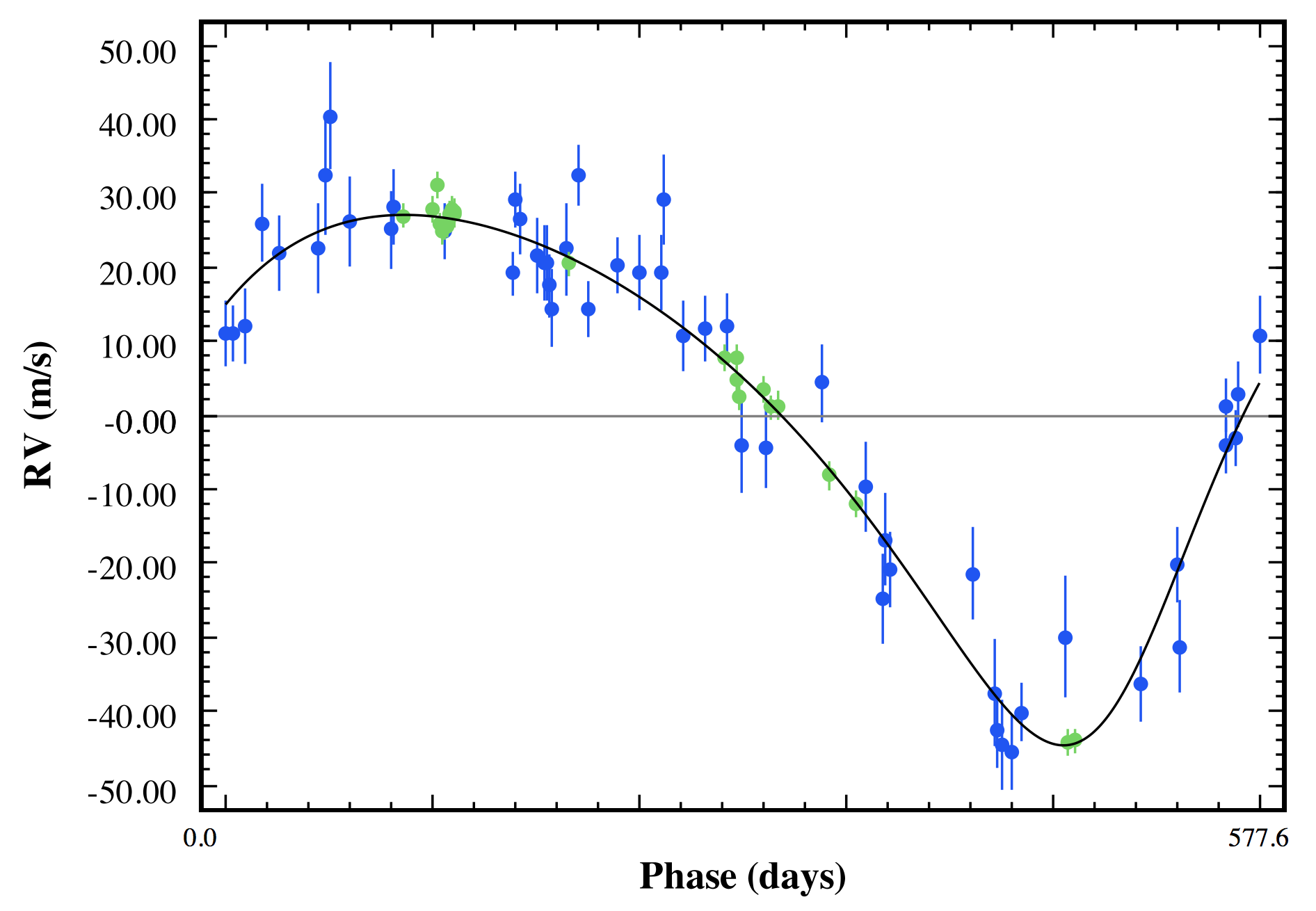}
    \includegraphics[width=0.65\columnwidth]{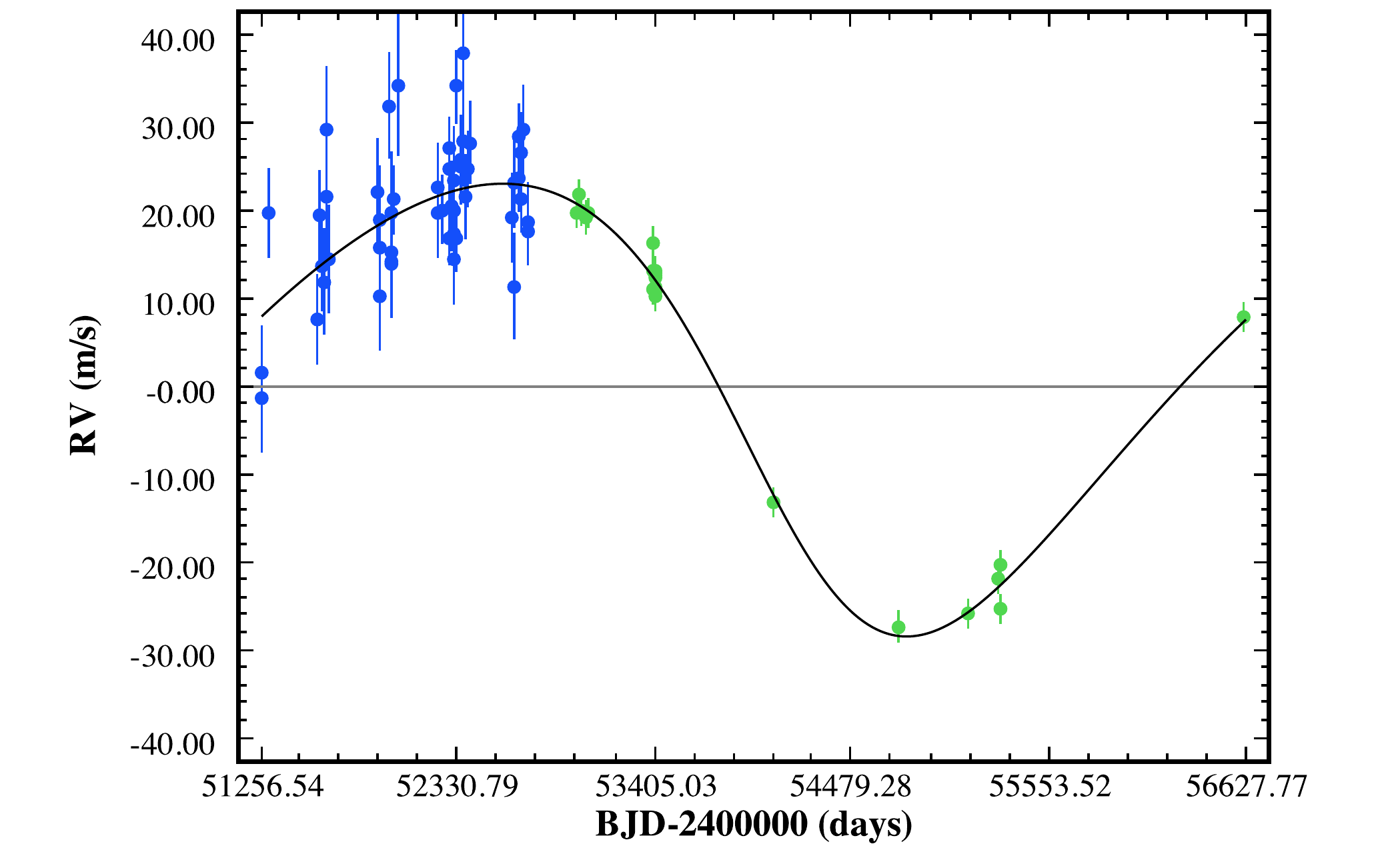}
    \includegraphics[width=0.65\columnwidth]{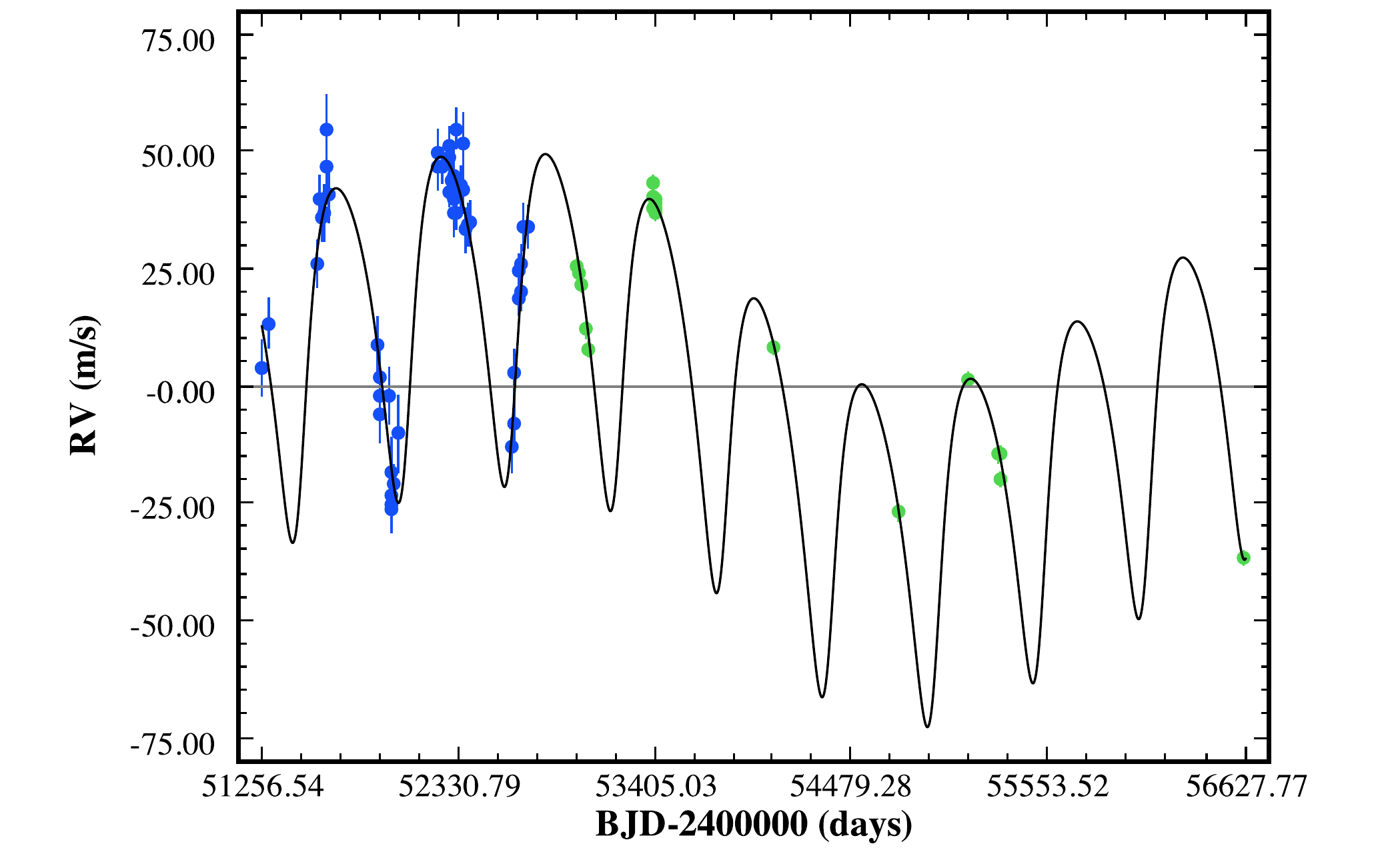}
    \caption{Left: Phase-folded data and model fit for HD\,65216b, with the outer planet removed.  Centre: Data and model fit for the outer planet HD\,65216c.  The signal of the inner planet has been removed.  Right: Data and model fit for both planets. The colours have the same meaning as in Figure~\ref{fig:7449}. }
    \label{fig:65216}
\end{figure*}

This system was originally reported to host a single eccentric planet with $P=613$ days and $e=0.41$ \citep{mayor04}.  \citet{songhu} speculated that the system may be decomposed into two circular planets at 572 and 152 days.  Our fit now excludes the possibility of an inner planet, strongly favouring instead a Jupiter analogue with $P=14.8$\,yr at $e=0.17$.  We also find the shorter period for HD\,65216b proposed by \citet{songhu} is preferred.  The new 2-planet fit (Figure~\ref{fig:65216}) also reduces the eccentricity of the inner planet to $e=0.29\pm$0.03, similar to the case of the HD\,159868 system \citep{witt12}.  It is possible for the signal of a long-period, low-amplitude planet to be mimicked by a stellar magnetic activity cycle, which for solar-type stars targeted by RV surveys has a typical duration $\sim$5-15 years and amplitudes of up to $\sim$10\,\ms\ \citep{if10, lovis11, yee18}.  With a large and well-constrained velocity semi-amplitude of $K_c=25.8\pm$1.3\,\ms, we are confident that the signal seen for HD\,65216 is due to an orbiting planet and not a stellar magnetic cycle.  A check of 9.0 years of ASAS photometry for HD\,65216 \citep{asas2} shows no significant periodicities, with a mean of 7.960$\pm$0.012 mag.  Furthermore, examination of the HARPS line bisector inverse span (BIS) yields no significant correlation with the RVs, as determined by a Pearson's correlation test.

\subsection{HD 85390}

\begin{figure}
	\includegraphics[width=\columnwidth]{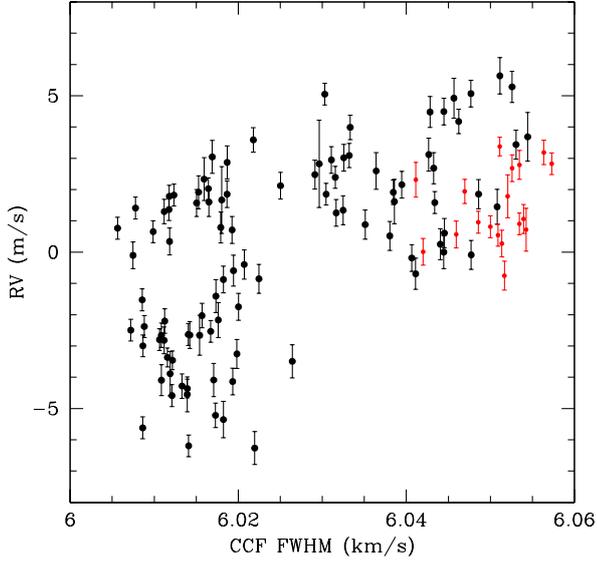}
    \caption{Residual RV for HD\,85390 (after removing the signal of HD\,85390b) as a function of the HARPS CCF FWHM. The clear correlation leads us to conclude that the residual velocity signal is activity-induced and not due to a long-period planet.}
    \label{fig:nope}
\end{figure}

\begin{figure*}
	\includegraphics[width=0.6\columnwidth]{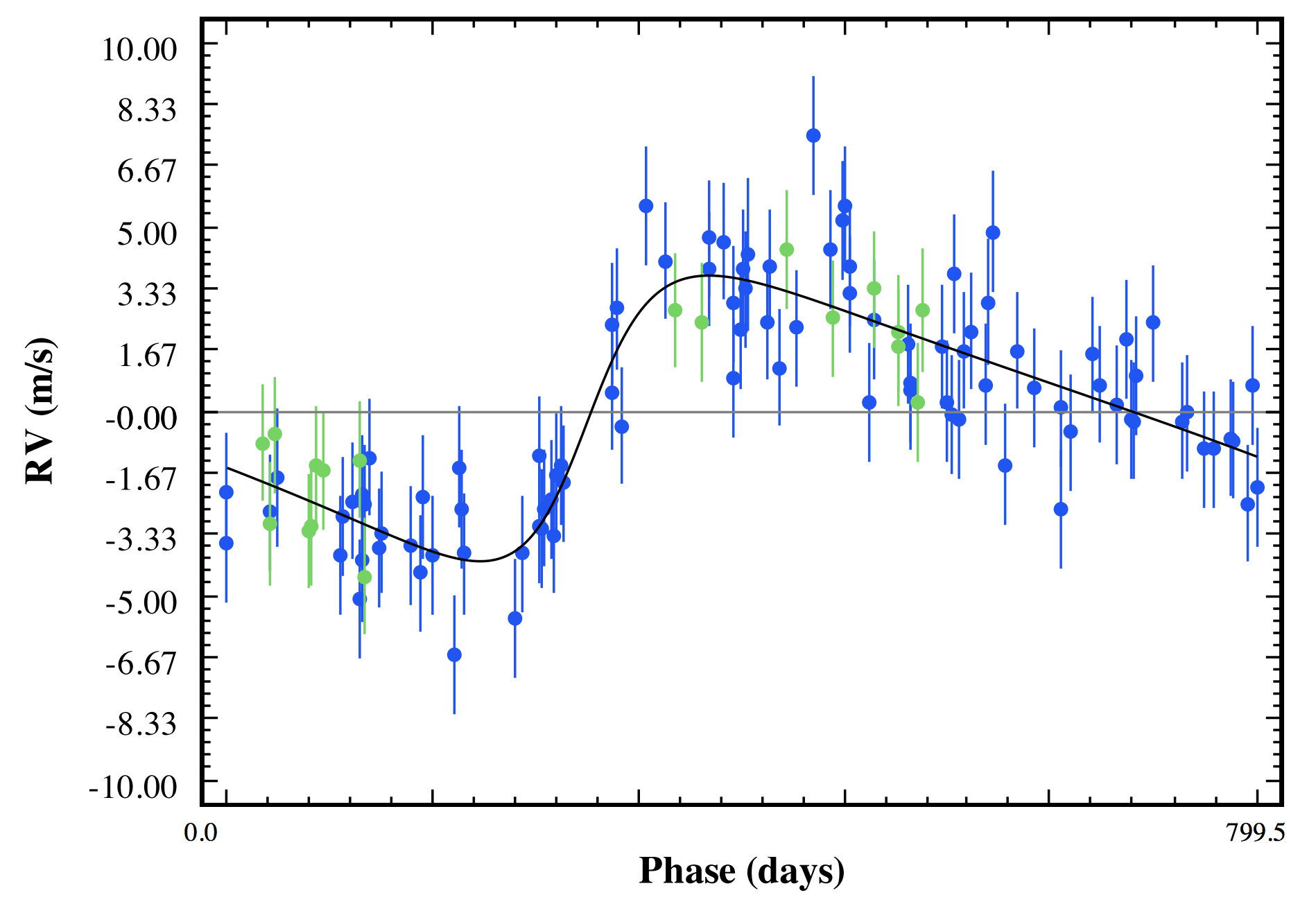}
    \includegraphics[width=0.65\columnwidth]{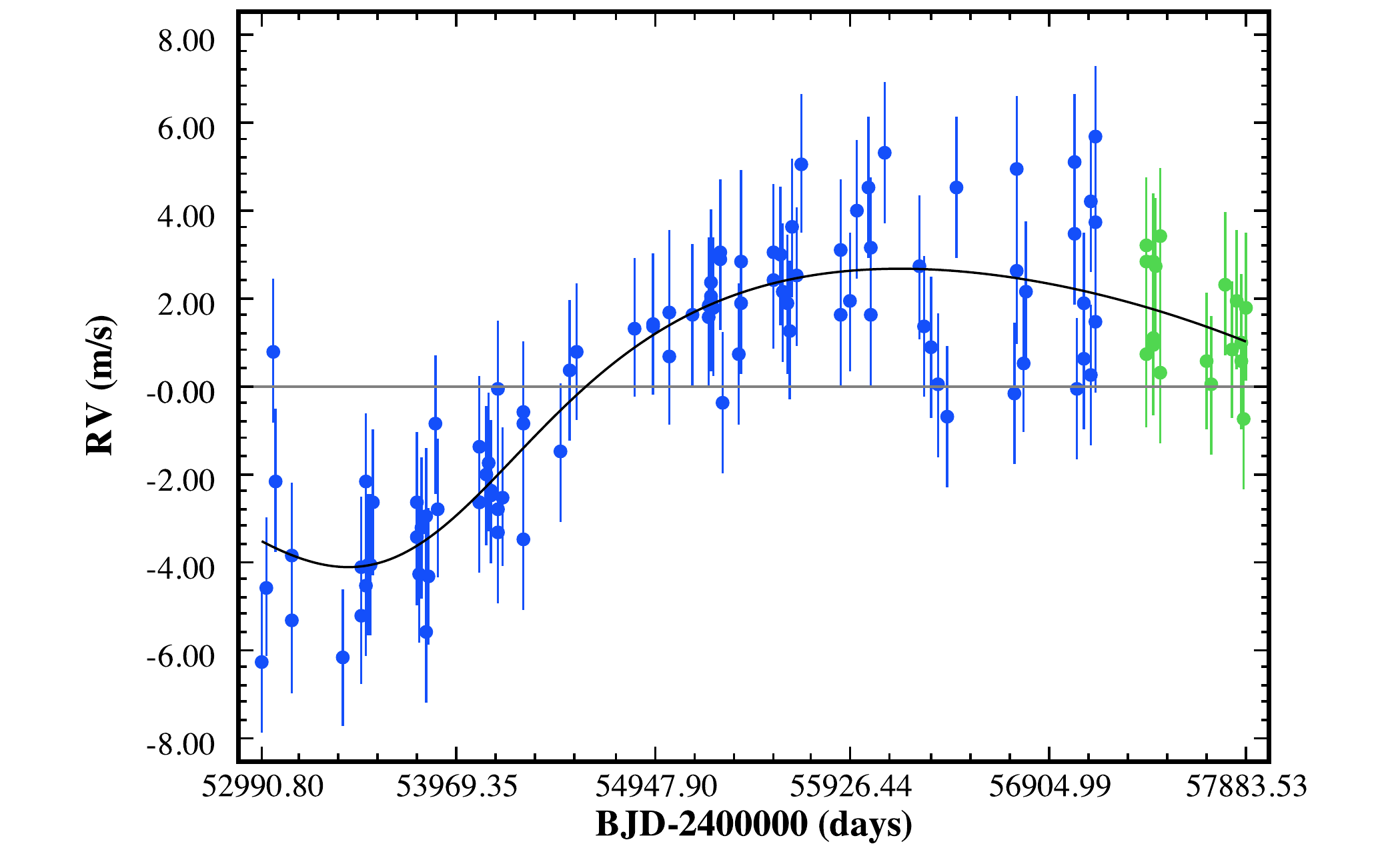}
    \includegraphics[width=0.65\columnwidth]{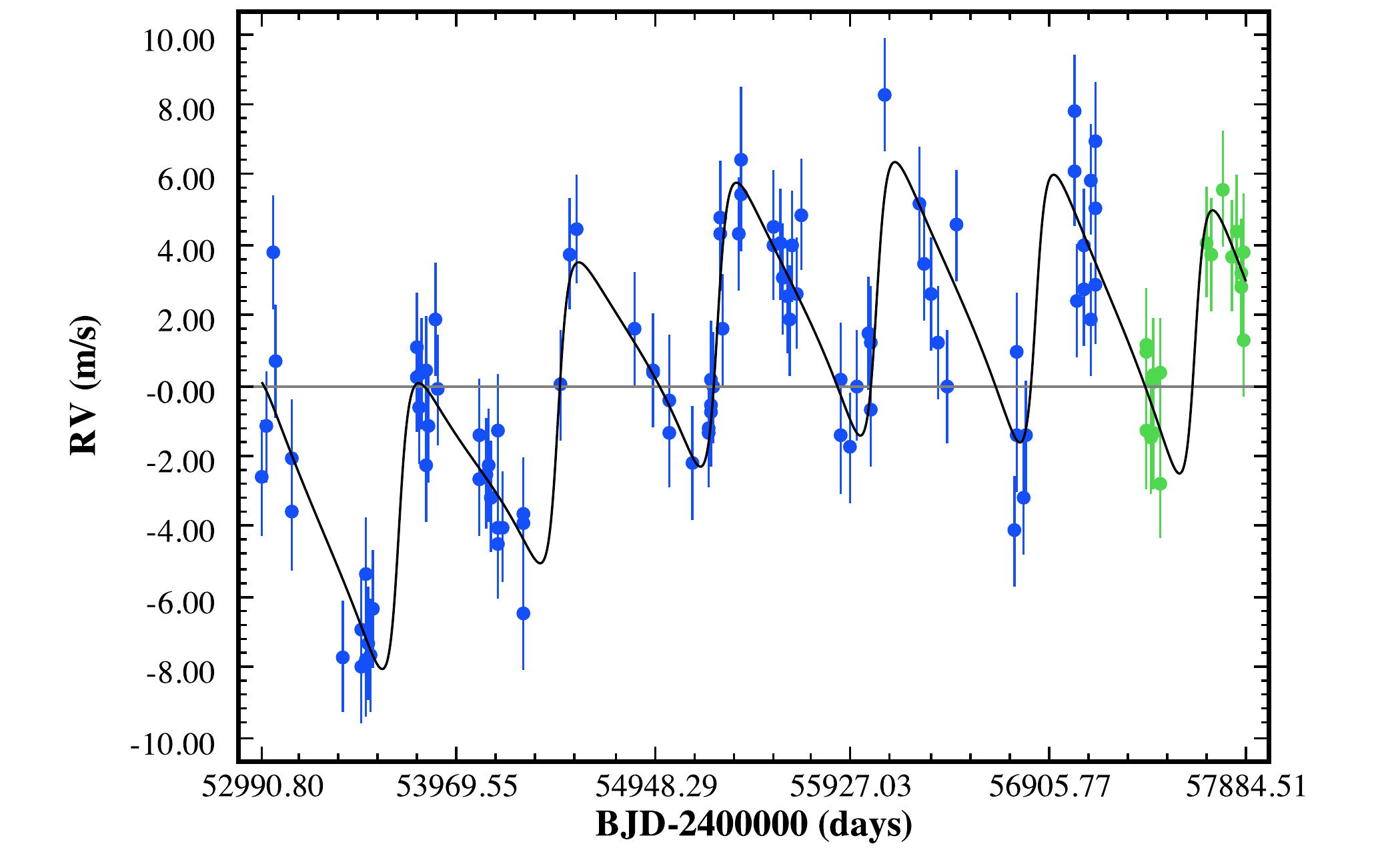}
    \caption{Left: Phase-folded data and model fit for HD\,85390b, with the secondary signal removed.  Centre: Data and model fit for the activity-induced variations, which have been modelled by a Keplerian orbit.  The signal of the inner planet has been removed.  Right: Data and model fit for both signals. The colours have the same meaning as in Figure~\ref{fig:7449}. }
    \label{fig:85390}
\end{figure*}

In the discovery work, \citet{mordasini11} noted an unconstrained long-period companion in addition to the 788-day planet HD\,85390b.  \citet{songhu} proposed a solution featuring two circular planets, at periods of 822 and 3700 days, which improved the fit to the extant HARPS velocities.  We now find that the long-period variation can be fit with a Keplerian orbit with a period of 18.3 years and an amplitude of 3.4$\pm$0.4\,\ms.  However, noting the small amplitude and fearing a stellar magnetic cycle, we checked the FWHM of the CCF as reported in the HARPS data headers.  Figure~\ref{fig:nope} shows the residual RV (after removing the securely-detected 790-day planet) as a function of the CCF FWHM.  There is a clear correlation, with a Pearson correlation coefficient of 0.6074, significant at more than 99.9\%.\footnote{Pearson R Calculator. (2019 Jan 17) Retrieved from \url{https://www.socscistatistics.com/pvalues/pearsondistribution.aspx }}  We thus conclude that the long-period signal is activity-induced.  This result highlights the need for vigilance as long-period, low-amplitude signals emerge from RV surveys \citep{endl16}.  The parameters for HD\,85390b are given in Table~\ref{tab:Bayesfit} with a ``planetary'' signal fit and removed to account for the long-period variation (Figure~\ref{fig:85390}).  The parameters of HD\,85390b are consistent with \citet{mordasini11}, and we find no evidence for further companions. 


\subsection{HD 89744}

\begin{figure*}
	\includegraphics[width=0.92\columnwidth]{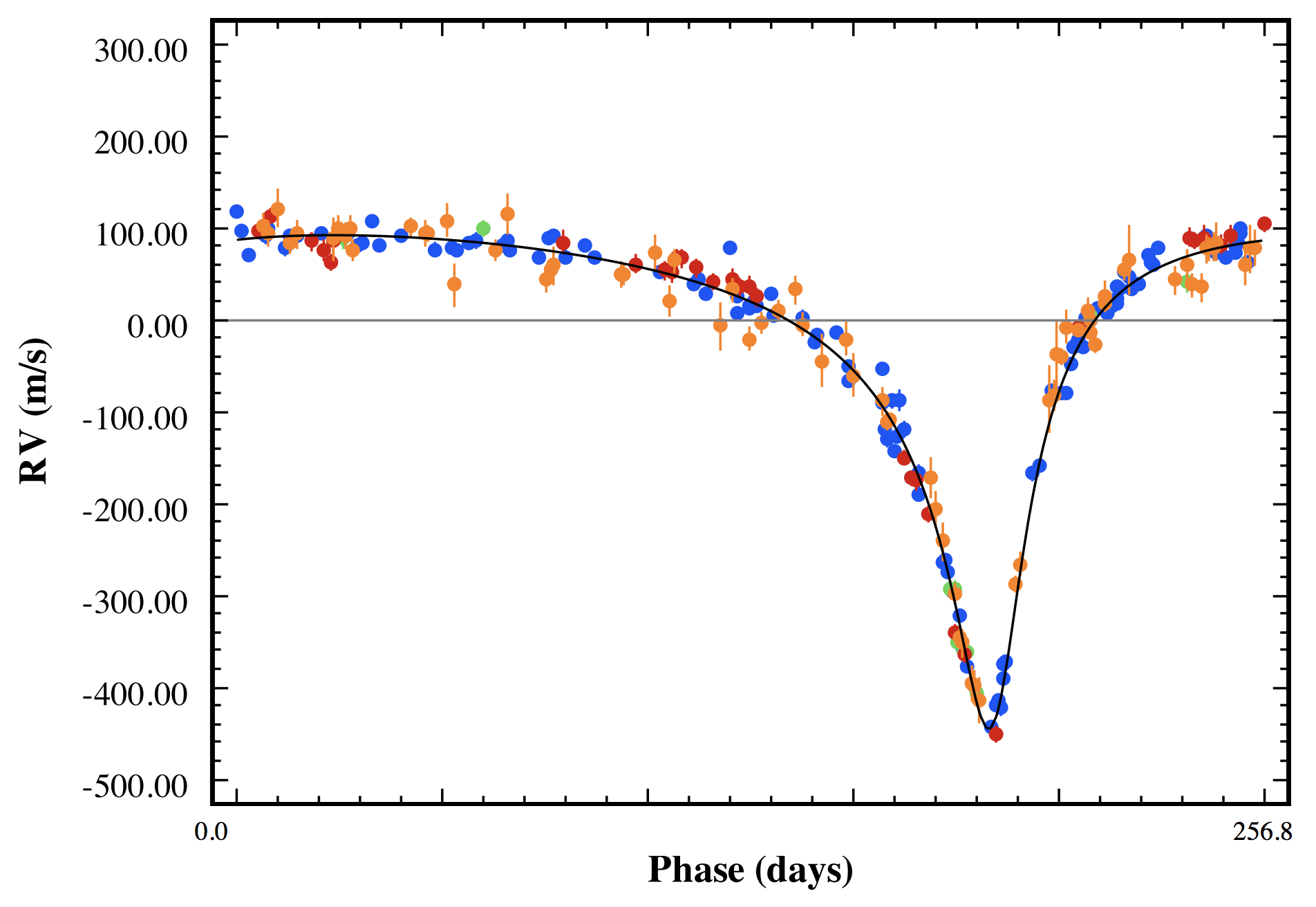}
    \includegraphics[width=\columnwidth]{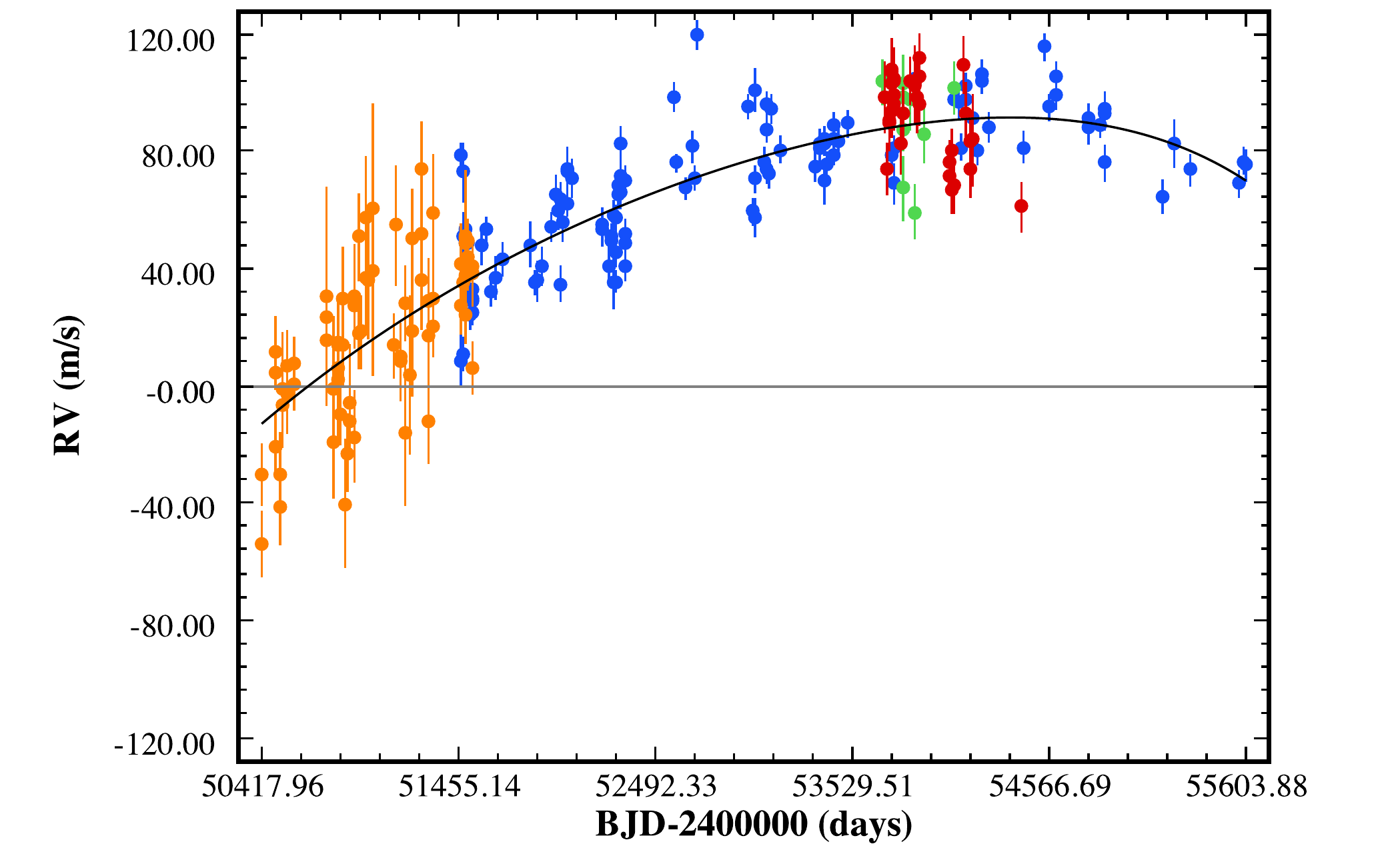}
    \caption{Left: Phase-folded fit for HD\,89744b.  Right: Data and model fit for the outer body HD\,89744c.  The signal of the inner planet has been removed. Orange: AFOE, blue: Lick, green: HJS, red: HET. }
    \label{fig:89744}
\end{figure*}

We fit four data sets for this system.  Much as for HD\,7449b, we find that the high eccentricity of HD\,89744b is solidly supported, and no near-circular double solutions make sense.  However, we find a long-period companion whose orbit appears to have turned over (Figure~\ref{fig:89744}).  Adding this signal reduces the rms from 23.8 to 16.4 m/s.  The outer companion is about 5 Jupiter masses and has an orbital period of about 19 years.  The amplitude and period of the outer body are much larger than would be expected for an activity-induced velocity variation.  The only publicly-available photometry is from \textit{Hipparcos} \citep{hipp}, and no periodicities are evident in the three-year time series. 


\subsection{HD 92788}

\begin{figure*}
	\includegraphics[width=0.92\columnwidth]{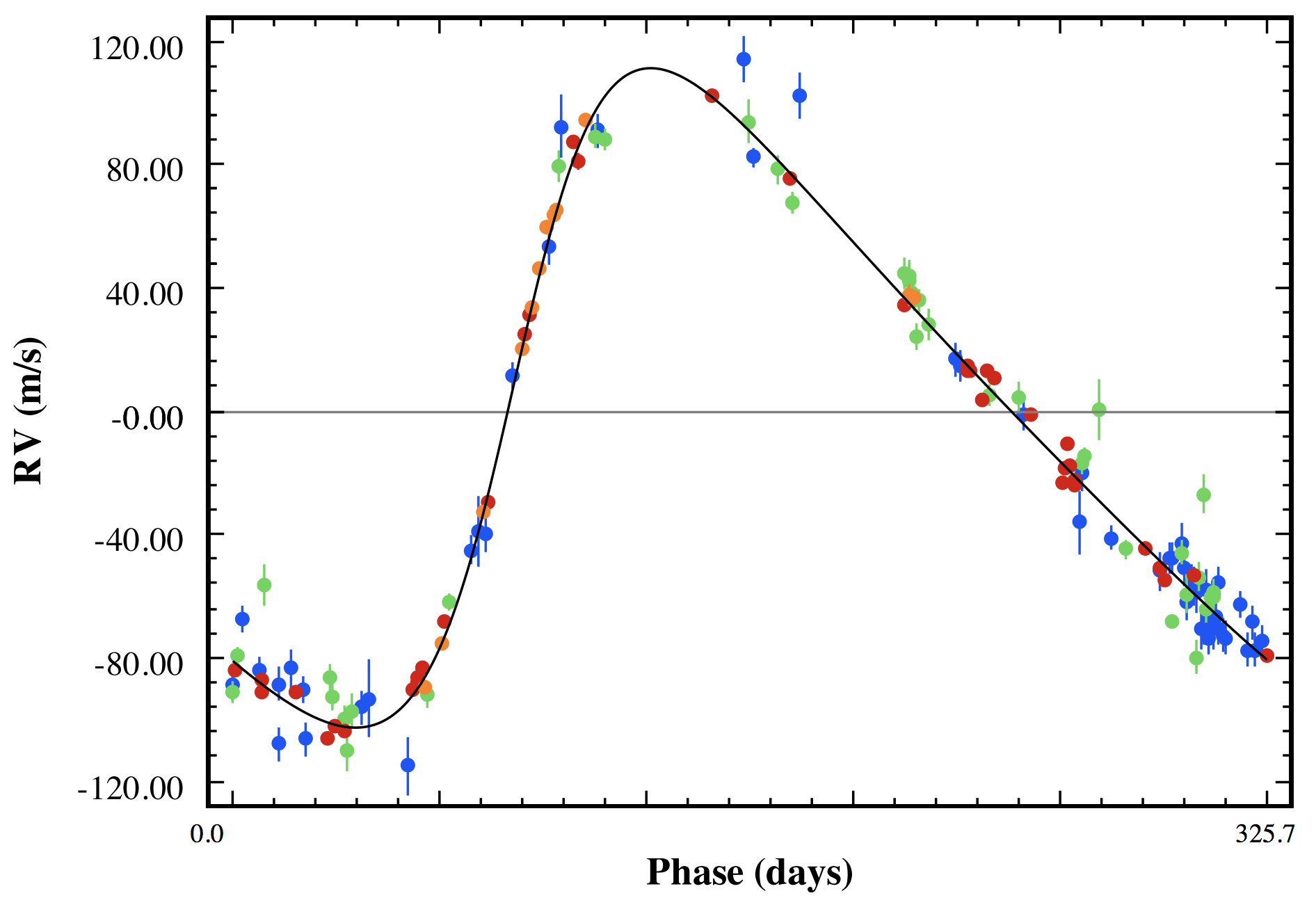}
    \includegraphics[width=\columnwidth]{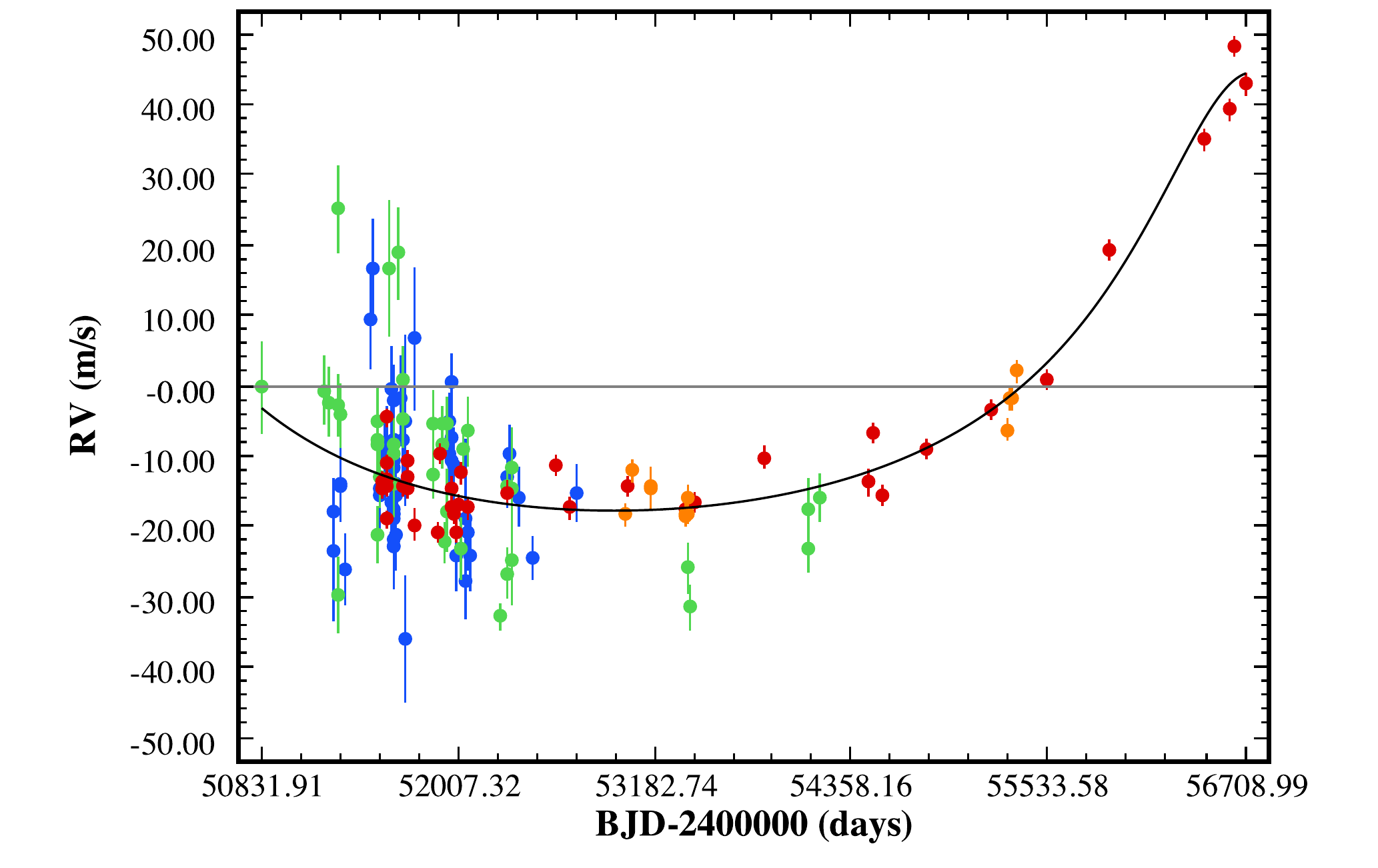}
    \caption{Left: Phase-folded fit for HD\,92788b.  Right: Data and model fit for the outer body HD\,92788c.  The signal of the inner planet has been removed.  Orange: HARPS, blue: CORALIE, green: Lick, red: Keck. }
    \label{fig:92788}
\end{figure*}

We have four data sets for HD\,92788, from CORALIE, HARPS, Lick, and Keck. As for HD\,7449 and HD\,89744, we find an unconstrained long-period signal that can be fit as a planet with 2.9\,\Mjup, though of course the period and mass remain poorly constrained due to an insufficient baseline of observations.  Analysis of 7.7 years of ASAS photometry \citep{asas2} with a mean of 7.316$\pm$0.019 yields no significant periodicities.  The fits are plotted in Figure~\ref{fig:92788}.  

\subsection{HD 117618}

As with HD\,3651 and HD\,52265, this system shows no indication of additional planets, and the additional unpublished AAT observations (Table~\ref{tab:117618vels}) continue to support the moderately eccentric single-planet system obtained by \citet{tinney05}, now achieving a total rms of 6.16\,\ms.

\begin{table*}
	\centering
	\caption{Results from \textit{astroEMPEROR} Fits}
	\label{tab:Bayesfit}
	\begin{tabular}{llllllll} 
		\hline
		Planet & Period & Eccentricity & $\omega$ & $\phi$ & $K$ & m sin $i$ & $a$ \\
           &  days &  & degrees & degrees & m/s & \Mjup & au \\
		\hline  

        HD 3651b & 62.250$\pm$0.004 & 0.645$\pm$0.02 &  243$\pm$3 & 140$\pm$3 & 16.6$\pm$0.6 & 0.228$\pm$0.011 & 0.295$\pm$0.029 \\       
        HD 7449b & 1255.5$\pm$5.1 & 0.92$\pm$0.03 & 69.9$\pm$5.7 & 267.6$\pm$ 1.2 & 21.9$\pm$2.6 & 0.508$\pm$0.111 & 2.38$\pm$0.04  \\
        HD 7449c & 15441$\pm$1059 & 0.0 (fixed) & 50$\pm$16* & 87$\pm$16 & 144$\pm$31 & 19.2$\pm$4.2 & 12.7$\pm$0.6 \\
		HD 52265b & 119.27$\pm$0.02 & 0.27$\pm$0.02 & 242$\pm$3 & 18$\pm$3 & 42.97$\pm$0.70 & 1.21$\pm$0.05 & 0.520$\pm$0.009 \\       
        HD 65216b & 577.6$\pm$1.328 & 0.27$\pm$0.02 & 212$\pm$3 & 238$\pm$3 & 35.70$\pm$1.28 & 1.295$\pm$0.062 & 1.301$\pm$0.020 \\
        HD 65216c & 5370$\pm$20 & 0.17$\pm$0.04 & 123$\pm$10 & 167$\pm$9 & 26.0$\pm$1.2 & 2.03$\pm$0.11 & 5.75$\pm$0.09 \\
        HD 85390b & 799.52$\pm$2.41 & 0.50$\pm$0.05 & 250$\pm$8 & 76$\pm$6 & 3.8$\pm$0.3 & 0.099$\pm$0.010 & 1.373$\pm$0.035 \\    
        HD 89744b & 256.78$\pm$0.02 & 0.677$\pm$0.003 & 193.7$\pm$0.4 & 275.46$\pm$0.24 & 269.66$\pm$1.45 & 8.35$\pm$0.18 & 0.917$\pm$0.009 \\
        HD 89744c & 6974$\pm$2161 & 0.29$\pm$0.12 & 174$\pm$26 & 65$\pm$32 & 45.1$\pm$38.5 & 5.36$\pm$4.57 & 8.3$\pm$1.8 \\
        HD 92788b & 325.72$\pm$0.03 & 0.351$\pm$0.004 & 277.6$\pm$0.6 & 78.2$\pm$0.4 & 108.5$\pm$0.6 & 3.78$\pm$0.18 & 0.971$\pm$0.023 \\
        HD 92788c & 9857$\pm$926 & 0.18$\pm$0.08 & 262$\pm$27 & 150$\pm$27 & 32.2$\pm$5.7 & 3.64$\pm$0.69 & 9.43$\pm$0.63 \\
        HD 117618b & 25.80$\pm$0.004 & 0.15$\pm$0.07 & 289$\pm$64 & 278$\pm$50 & 10.90$\pm$0.68 & 0.174$\pm$0.014 & 0.180$\pm$0.005 \\
    
        \hline
	\end{tabular}
\end{table*}

\begin{table*}
	\centering
	\caption{Jitter and Offset values for \textit{astroEMPEROR} Fits}
	\label{tab:Bayesjit}
	\begin{tabular}{lllllllll} 
    
    \hline
    Star & $J_0$ & $O_0$ & $J_1$ & $O_1$ & $J_2$ & $O_2$ & $J_3$ & $O_3$ \\
     & m/s & m/s & m/s & m/s & m/s & m/s & m/s & m/s \\
    \hline
    
    HD3651 &  4.30$\pm$0.73$^a$ & 1.29$\pm$0.79$^a$ & 1.55$\pm$0.49$^b$ & -10.02$\pm$2.37$^b$ & 3.00$\pm$0.23$^c$ & 0.92$\pm$0.29$^c$ & 6.97$\pm$0.37$^d$ & 0.42$\pm$0.52$^d$ \\
    HD7449 & 8.73$\pm$0.52$^e$ & 85.51$\pm$19.12$^e$ & 4.08$\pm$0.50$^{e*}$ & -9.96$\pm$21.48$^{e*}$ \\
    HD5226 & 7.38$\pm$1.13$^f$ & 0.91$\pm$0.99$^f$ & 3.44$\pm$1.65$^e$ & -21.99$\pm$2.23$^e$ & 4.09$\pm$0.38$^c$ & -1.69$\pm$0.82$^c$ \\
    HD65216 & 2.84$\pm$0.87$^f$ & -24.39$\pm$1.95$^f$ & 0.27$\pm$0.35$^e$ & -15.40$\pm$3.13$^e$ \\
    HD85390 & 0.73$\pm$0.27$^e$ & -3.54$\pm$0.23$^e$ & 0.06$\pm$0.30$^{e*}$ & -8.11$\pm$0.33$^{e*}$ \\
    HD89744 & 11.54$\pm$1.56$^g$ & -5.40$\pm$43.13$^g$ & 10.18$\pm$0.86$^a$ & -93.88$\pm$43.22$^a$ & 8.88$\pm$1.32$^b$ & 65.80$\pm$43.15$^b$ & 12.66$\pm$0.74$^d$ & -48.20$\pm$43.11$^d$ \\
    HD92788 & 6.37$\pm$0.54$^f$ & 35.18$\pm$8.37$^f$ & 1.40$\pm$0.51$^e$ & -30.94$\pm$8.31$^e$ & 3.57$\pm$0.39$^c$ & 13.67$\pm$8.31$^c$ & 10.23$\pm$0.95$^d$ & 12.50$\pm$8.39$^d$ \\
    HD117618 & 5.46$\pm$0.45$^h$ & -2.87$\pm$1.02$^h$ & 1.57$\pm$0.80$^e$ & 1.42$\pm$1.34$^e$ & 6.55$\pm$0.84$^{e*}$ & -6.87$\pm$2.80$^{e*}$ \\
    \hline
    \multicolumn{9}{l}{$^a$HET/HRS $^b$2.7m/CS23 $^c$Keck/HIRES $^d$Lick $^e$HARPS $^f$CORALIE  $^g$AFOE $^h$AAT/UCLES}
    \\
	\end{tabular}
\end{table*}


\clearpage

\begin{table}
	\centering
	\caption{\textit{astroEMPEROR} fit properties for candidate multiple systems}
	\label{tab:bicmeasuring2}
	\begin{tabular}{llllllll} 
		\hline
		Star & $N_{planets}$ & BIC & $\Delta$BIC & AIC & $\Delta$AIC & MAP & $\Delta$MAP \\
		\hline
		HD 7449 	& 2 & 872.7 & 0.0 & 914.1 & 0.0 & -421.3 & 0.0 \\
                	& 1 & 981.0 & 108.3 & 953.3 & 39.2 & -466.7 & 45.4 \\
		HD 65216 	& 2 & 486.1 & 0.0 & 451.1 & 0.0 & -210.6 & 0.0 \\
                	& 1 & 589.4 & 103.3 & 566.1 & 115 & -273.1 & 62.5 \\
		HD 89744 	& 2 & 2051.5 & 0.0 & 1985.9 & 0.0 & -974.0 & 0.0 \\
                	& 1 & 2106.2 & 54.7 & 2057.9 & 72.0 & -1014.9 & 40.9 \\
		HD 92788 	& 2 & 1051.0 & 0.0 &  994.6 & 0.0 & -478.3 & 0.0 \\
        			& 1 & 1137.4 & 86.4 & 1095.9 & 101.3 & -534.0 & 55.7 \\
\hline
	\end{tabular}
\end{table}

\clearpage

\section{Discussion and Conclusions}
\label{Conclusions}
In this work, we set out to test the hypothesis put forward in \citet{songhu} that some moderately eccentric single-planet systems may be better fit with two low-eccentricity planets.  We gathered newly available RV data from the Lick and Keck major data releases \citep{fischer14,butler17}, as well as updated HARPS velocities from the ESO Archive.  

We find four long period candidate companions in systems where an eccentric planet is present.  If confirmed, these bodies could be responsible for eccentricity excitation via the Lidov-Kozai mechanism \citep{lidov, kozai}.  In particular, HD\,7449b and HD\,89744b are highly eccentric planets ($e>0.67$) which we confirm to be genuinely eccentric singles.  The presence of distant massive companions driving Kozai oscillations would explain the origin of the large eccentricities, though we note that the conditions required for such oscillations are stringent, e.g. the perturbing body must be inclined at $i\gtsimeq$39\degrees\ with respect to the orbital plane of the planet(s). 

In Wittenmyer et al. (2019a, submitted), we found that 'deceptive couplets' - two near-resonant planets moving on near-circular orbits - would most often masquerade as a single planet with orbital eccentricity of approximately $e=0.31\pm$0.12 in sparsely sampled data.  We also found that such systems were highly unlikely to imitate single planets with orbital eccentricities greater than $e\sim 0.5$.  In light of those results, it is interesting to note that the orbital parameters of the most eccentric of the systems studied in this work remain robust in light of the new data considered.  

Eccentricity and multiplicity are two of the most important factors to understand exoplanetary dynamical history.  The precise distributions of them, however, are hard to measure.

On the one hand, eccentricity is one of most poorly constrained orbital parameters, for two primary reasons:
1). Precise measurements of eccentricity require intense phase coverage, which is very expensive, and thus not valued by the most RV surveys.
2). Eccentricity is very sensitive to correlated noise, which is very common in RV data. The moderate eccentricity ($e=0.2$) for GJ\,876d \citep{Rivera2010}, a super-Earth with period of about 2~days, is a long-standing puzzle in the field.  A recent analysis confirmed its circular orbit, and proved the previously reported eccentricity is most likely caused by the correlated noise \citep{Millholland2018}.  It has previously been noted that Keplerian fits to RV data tend to be biased against $e\sim$0 \citep{shen08,otoole09}. 

At the same time, it remains hugely challenging to precisely determine the number of planets in systems monitored by RV planet search programs. In addition to degeneracies such as that described in this work, there is a clear floor below which planets could exist, but remain undetected. As new technology and new instrumentation comes online, that floor can be suppressed, revealing previously hidden planets.


Given the challenges involved in discovering them, and searching for additional planets within, the true multiplicity of exoplanetary systems remains even more controversial. To give but one example - most systems that contain hot Jupiters appear to only host a single, isolated planet \citep{Steffen2012, wang2018a, wyh2017, wxy2018}, which many have suggested could be the result of the migration required to move such a planet in to the brink of its host star's atmosphere. Though this is clearly plausible, the apparent isolated nature of hot Jupiters could also be the result of detection bias \citep{becker2015, Millholland2016}.  In general, however, one fact remains true: the closer we look, the more planets are found.

The challenges involved in finding hidden worlds are particularly apparent for those systems studied by RV surveys, where there is a clear and well established degeneracy between eccentricity and multiplicity, especially for planets close to or in mean motion resonance.  They can easily confound each other \citep{trifonov17}.  In this light, our work is extremely important to determine the true configuration of exoplanetary systems.  Our results suggest that a subgroup of warm Jupiters may be delivered by the Kozai-Lidov mechanism \citep{Wu2003}.  We also did not detect any additional companion for a circular warm Jupiter system (HD\,117618).  It can be explained as the result of \textit{in situ} formation \citep{Batyginetal2016}.  Additional RV follow-up is urgently needed to search for additional close-in low-mass companions in the system, which is the natural prediction from \textit{in situ} formation.  The possibility of such close-in low-mass companions in systems containing cool giant planets was explored by \citep{witt09}, who monitored 22 systems but found no interior planets to a limit of about 10-15\,Mearth. 


Interestingly, for the remaining four eccentric cold Jupiters examined here (HD\,7449b, HD\,65216b, HD\,89744b, and HD\,92788b), we found evidence for long-term substellar companions with minimum masses between $2-19\,$\Mjup.  Although a massive perturber can produce moderate eccentricity \citep{Anderson2017}, the existence of extremely high eccentricity (e.g. $e=0.92$ for HD 7449) suggests that planet-planet scattering did occur.  But it is still unclear which channel is the dominant mechanism for producing short-period gas giants (see \citealt{Dawson2018} for review, and see also \citealt{wang2018b, wang2018c}), because we do not have the 3-D orbital configuration.  The ongoing {\it GAIA} mission will shed new light on distinguishing planet-planet interaction \citep{Ford2008} and the Lidov-Kozai mechanism \citep{Wu2003} by providing 3-D orbital information. 

In sum, our results highlight the importance of revisiting the analysis of confirmed exoplanetary systems once additional data becomes available.  They also reinforce the critical need for legacy exoplanet surveys (such as AAPS) to continue, where possible, obtaining data on their target stars - or for new RV surveys (such as {\sc Minerva}-Australis, \citealt{witt18,brett}) to include such stars in their observing schedules, as targets for occasional observational follow-up.

\section*{Acknowledgements}
S.W. thanks the Heising-Simons Foundation for their generous support.
DJ acknowledges support from the National Science Foundation under Grant No. 1559487 and 1559505. JC's research is supported by an Australian Government Research Training Program (RTP) Scholarship. We acknowledge the traditional owners of the land on which the AAT stands, the Gamilaraay people, and pay our respects to elders past, present and emerging.  This work made use of publicly-available HARPS spectra from the ESO Archive, with the following program IDs: 192.C-0852(A), 183.C-0972(A), 072.C-0488(E), 097.C-0090(A), 091.C-0936(A), 198.C-0836(A), 089.D-0302(A), 60.A-9036(A), 073.D-0578(A), 60.A-9700(G).





\begin{thebibliography}{99}

\bibitem[Anderson \& Lai(2017)]{Anderson2017} Anderson, K.~R., \& Lai, D.\ 2017, \mnras, 472, 3692 

\bibitem[Anglada-Escud{\'e} et al.(2010)]{ang10} Anglada-Escud{\'e}, G., L{\'o}pez-Morales, M., \& Chambers, J.~E.\ 2010, \apj, 709, 168 

\bibitem[Addison et al.(2019)]{brett} Addison, B., Wright, D., Wittenmyer, R.~A, et al.\ 2019, \pasp, submitted 

\bibitem[Batygin \& Brown(2016)]{Batygin2016} Batygin, K., \& Brown, M.~E.\ 2016, \aj, 151, 22 

\bibitem[Batygin et al.(2016)]{Batyginetal2016} Batygin, K., Bodenheimer, P.~H., \& Laughlin, G.~P.\ 2016, \apj, 829, 114 

\bibitem[Becker et al.(2015)]{becker2015} Becker, J.~C., Vanderburg, A., Adams, F.~C., Rappaport, S.~A., \& Schwengeler, H.~M.\ 2015, \apjl, 812, L18

\bibitem[Boisvert et al.(2018)]{boisvert18} Boisvert, J.~H., Nelson, B.~E., \& Steffen, J.~H.\ 2018, \mnras, 480, 2846 

\bibitem[Bonfils et al.(2013)]{bonfils13} Bonfils, X., Delfosse, X., Udry, S., et al.\ 2013, \aap, 549, A109 

\bibitem[Bryan et al.(2018)]{bryan18} Bryan, M.~L., Knutson, H.~A., Fulton, B., et al.\ 2018, arXiv:1806.08799 

\bibitem[Butler \& Marcy(1996)]{early1} Butler, R.~P., \& Marcy, G.~W.\ 1996, \apjl, 464, L153 

\bibitem[Butler et al.(2017)]{butler17} Butler, R.~P., Vogt, S.~S., Laughlin, G., et al.\ 2017, \aj, 153, 208 

\bibitem[Campbell et al.(1988)]{campbell88} Campbell, B., Walker, G.~A.~H., \& Yang, S.\ 1988, \apj, 331, 902 

\bibitem[Carter et al.(2003)]{carter03} Carter, B.~D., Butler, R.~P., Tinney, C.~G., et al.\ 2003, \apjl, 593, L43 

\bibitem[Davis et al.(1984)]{davis} Davis, M., Hut, P., \& Muller, R.~A.\ 1984, \nat, 308, 715.

\bibitem[Dawson \& Fabrycky(2010)]{df10} Dawson, R.~I., \& Fabrycky, D.~C.\ 2010, \apj, 722, 937 

\bibitem[Dawson \& Johnson(2018)]{Dawson2018} Dawson, R.~I., \& Johnson, J.~A.\ 2018, \araa, 56, 175

\bibitem[D{\'{\i}}az et al.(2018)]{diaz18} D{\'{\i}}az, M.~R., Jenkins, J.~S., Tuomi, M., et al.\ 2018, \aj, 155, 126 

\bibitem[Dumusque et al.(2011)]{dum11} Dumusque, X., Lovis, C., S{\'e}gransan, D., et al.\ 2011, \aap, 535, A55 

\bibitem[Endl et al.(2016)]{endl16} Endl, M., Brugamyer, E.~J., Cochran, W.~D., et al.\ 2016, \apj, 818, 34 

\bibitem[Fischer et al.(2014)]{fischer14} Fischer, D.~A., Marcy, G.~W., \& Spronck, J.~F.~P.\ 2014, \apjs, 210, 5 

\bibitem[Fischer et al.(2016)]{fischer16} Fischer, D.~A., Anglada-Escude, G., Arriagada, P., et al.\ 2016, \pasp, 128, 066001 

\bibitem[Foreman-Mackey et al.(2013)]{fm13} Foreman-Mackey, D., Hogg, D.~W., Lang, D., \& Goodman, J.\ 2013, \pasp, 125, 306 

\bibitem[Ford \& Rasio(2008)]{Ford2008} Ford, E.~B., \& Rasio, F.~A.\ 2008, \apj, 686, 621 

\bibitem[Fressin et al.(2013)]{fressin13} Fressin, F., Torres, G., Charbonneau, D., et al.\ 2013, \apj, 766, 81 

\bibitem[Gregory(2005)]{g05} Gregory, P.~C.\ 2005, \apj, 631, 1198 

\bibitem[Hatzes et al.(2018)]{hatzes18} Hatzes, A.~P., Endl, M., Cochran, W.~D., et al.\ 2018, \aj, 155, 120 

\bibitem[Horner, \& Evans(2002)]{horner02} Horner, J., \& Evans, N.~W.\ 2002, \mnras, 335, 641.

\bibitem[Isaacson \& Fischer(2010)]{if10} Isaacson, H., \& Fischer, D.\ 2010, \apj, 725, 875 

\bibitem[Jenkins \& Tuomi(2014)]{jenkins14} Jenkins, J.~S., \& Tuomi, M.\ 2014, \apj, 794, 110 

\bibitem[Jones et al.(2010)]{jones10} Jones, H.~R.~A., Butler, R.~P., Tinney, C.~G., et al.\ 2010, \mnras, 403, 1703 

\bibitem[Korzennik et al.(2000)]{korzennik00} Korzennik, S.~G., Brown, T.~M., Fischer, D.~A., Nisenson, P., \& Noyes, R.~W.\ 2000, \apjl, 533, L147 

\bibitem[Kozai(1962)]{kozai} Kozai, Y.\ 1962, \aj, 67, 591 

\bibitem[K{\"u}rster et al.(2015)]{kurster15} K{\"u}rster, M., Trifonov, T., Reffert, S., Kostogryz, N.~M., \& Rodler, F.\ 2015, \aap, 577, A103 

\bibitem[Latham et al.(1989)]{lathamsworld} Latham, D.~W., Mazeh, T., Stefanik, R.~P., Mayor, M., \& Burki, G.\ 1989, \nat, 339, 38 

\bibitem[Lidov(1962)]{lidov} Lidov, M.~L.\ 1962, \planss, 9, 719 

\bibitem[Lo Curto et al.(2015)]{locurto15} Lo Curto, G., Pepe, F., Avila, G., et al.\ 2015, The Messenger, 162, 9 

\bibitem[Lovis et al.(2011)]{lovis11} Lovis, C., Dumusque, X., Santos, N.~C., et al.\ 2011, arXiv:1107.5325 

\bibitem[Lykawka, \& Mukai(2008)]{ss4} Lykawka, P.~S., \& Mukai, T.\ 2008, \aj, 135, 1161.

\bibitem[Matese, \& Whitmire(1986)]{ss1} Matese, J.~J., \& Whitmire, D.~P.\ 1986, \icarus, 65, 37.

\bibitem[Matese et al.(1999)]{ss2} Matese, J.~J., Whitman, P.~G., \& Whitmire, D.~P.\ 1999, \icarus, 141, 354.

\bibitem[Matese, \& Whitmire(2011)]{ss5} Matese, J.~J., \& Whitmire, D.~P.\ 2011, \icarus, 211, 926.

\bibitem[Mayor \& Queloz(1995)]{51peg} Mayor, M., \& Queloz, D.\ 1995, \nat, 378, 355 

\bibitem[Mayor et al.(2004)]{mayor04} Mayor, M., Udry, S., Naef, D., et al.\ 2004, \aap, 415, 391 

\bibitem[Meschiari et al.(2009)]{mes09} Meschiari, S., Wolf, A.~S., Rivera, E., et al.\ 2009, \pasp, 121, 1016 

\bibitem[Millholland et al.(2016)]{Millholland2016} Millholland, S., Wang, S., \& Laughlin, G.\ 2016, \apjl, 823, L7 

\bibitem[Millholland et al.(2018)]{Millholland2018} Millholland, S., Laughlin, G., Teske, J., et al.\ 2018, \aj, 155, 106 

\bibitem[Mordasini et al.(2011)]{mordasini11} Mordasini, C., Mayor, M., Udry, S., et al.\ 2011, \aap, 526, A111 

\bibitem[Murray(1999)]{ss3} Murray, J.~B.\ 1999, \mnras, 309, 31.

\bibitem[Naef et al.(2001)]{naef01} Naef, D., Mayor, M., Pepe, F., et al.\ 2001, \aap, 375, 205 

\bibitem[O'Toole et al.(2009)]{otoole09} O'Toole, S.~J., Tinney, C.~G., Jones, H.~R.~A., et al.\ 2009, \mnras, 392, 641 

\bibitem[Pojmanski(2002)]{asas1} Pojmanski, G.\ 2002, \actaa, 52, 397 

\bibitem[Pojmanski(2003)]{asas2} Pojmanski, G.\ 2003, \actaa, 53, 341 

\bibitem[Rajpaul et al.(2016)]{byebyeBb} Rajpaul, V., Aigrain, S., \& Roberts, S.\ 2016, \mnras, 456, L6 

\bibitem[Rivera et al.(2010)]{Rivera2010} Rivera, E.~J., Laughlin, G., Butler, R.~P., et al.\ 2010, \apj, 719, 890

\bibitem[Robertson et al.(2014)]{MurderDeathKill} Robertson, P., Mahadevan, S., Endl, M., \& Roy, A.\ 2014, Science, 345, 440 

\bibitem[Robertson et al.(2015)]{nokapteyn} Robertson, P., Roy, A., \& Mahadevan, S.\ 2015, \apjl, 805, L22 

\bibitem[Santos et al.(2016)]{santos16} Santos, N.~C., Santerne, A., Faria, J.~P., et al.\ 2016, \aap, 592, A13 

\bibitem[Shen \& Turner(2008)]{shen08} Shen, Y., \& Turner, E.~L.\ 2008, \apj, 685, 553 

\bibitem[Sousa et al.(2008)]{sousa08} Sousa, S.~G., Santos, N.~C., Mayor, M., et al.\ 2008, \aap, 487, 373 

\bibitem[Steffen et al.(2012)]{Steffen2012} Steffen, J.~H., Ragozzine, D., Fabrycky, D.~C., et al.\ 2012, Proceedings of the National Academy of Science, 109, 7982 

\bibitem[Tal-Or et al.(2019)]{talor18} Tal-Or, L., Trifonov, T., Zucker, S., Mazeh, T., \& Zechmeister, M.\ 2019, \mnras, 484, L8  

\bibitem[Tinney et al.(2005)]{tinney05} Tinney, C.~G., Butler, R.~P., Marcy, G.~W., et al.\ 2005, \apj, 623, 1171 

\bibitem[Tinney et al.(2011)]{tinney11} Tinney, C.~G., Wittenmyer, R.~A., Butler, R.~P., et al.\ 2011, \apj, 732, 31 

\bibitem[Trifonov et al.(2017)]{trifonov17} Trifonov, T., K{\"u}rster, M., Zechmeister, M., et al.\ 2017, \aap, 602, L8 

\bibitem[Valenti \& Fischer(2005)]{vf05} Valenti, J.~A., \& Fischer, D.~A.\ 2005, \apjs, 159, 141 

\bibitem[van Leeuwen(2007)]{hipp} van Leeuwen, F.\ 2007, \aap, 474, 653 

\bibitem[Wang et al.(2018a)]{wang2018a} Wang, S., Wang, X.-Y., Wang, Y.-H., et al.\ 2018c, \aj, 156, 181
  
\bibitem[Wang et al.(2018b)]{wang2018b} Wang, S., Addison, B., Fischer, D.~A., et al.\ 2018, \aj, 155, 70 

\bibitem[Wang et al.(2019)]{wang2018c} Wang, S., Jones, M., Shporer, A., et al.\ 2019, \aj, 157, 51 

\bibitem[Wang X. et al.(2018)]{wxy2018} Wang, X.-Y., Wang, S., Hinse, T.~C., et al.\ 2018, \pasp, 130, 064401

\bibitem[Wang Y. et al.(2017)]{wyh2017} Wang, Y.-H., Wang, S., Liu, H.-G., et al.\ 2017, \aj, 154, 49

\bibitem[Wittenmyer et al.(2009)]{witt09} Wittenmyer, R.~A., Endl, M., Cochran, W.~D., Levison, H.~F., \& Henry, G.~W.\ 2009, \apjs, 182, 97 

\bibitem[Wittenmyer et al.(2011)]{etaearth} Wittenmyer, R.~A., Tinney, C.~G., Butler, R.~P., et al.\ 2011, \apj, 738, 81 

\bibitem[Wittenmyer et al.(2012)]{witt12} Wittenmyer, R.~A., Horner, J., Tuomi, M., et al.\ 2012, \apj, 753, 169 

\bibitem[Wittenmyer et al.(2013a)]{songhu} Wittenmyer, R.~A., Wang, S., Horner, J., et al.\ 2013a, \apjs, 208, 2 

\bibitem[Wittenmyer et al.(2013b)]{witt13pasp} Wittenmyer, R.~A., Tinney, C.~G., Horner, J., et al.\ 2013b, \pasp, 125, 351 

\bibitem[Wittenmyer et al.(2014)]{2jupiters} Wittenmyer, R.~A., Horner, J., Tinney, C.~G., et al.\ 2014, \apj, 783, 103 

\bibitem[Wittenmyer et al.(2016)]{jupiters} Wittenmyer, R.~A., Butler, R.~P., Tinney, C.~G., et al.\ 2016, \apj, 819, 28 

\bibitem[Wittenmyer et al.(2017a)]{30177} Wittenmyer, R.~A., Horner, J., Mengel, M.~W., et al.\ 2017a, \aj, 153, 167 

\bibitem[Wittenmyer et al.(2017b)]{76920} Wittenmyer, R.~A., Jones, M.~I., Horner, J., et al.\ 2017b, \aj, 154, 274 

\bibitem[Wittenmyer et al.(2018)]{witt18} Wittenmyer, R.~A, Horner, J., Carter, B.~D, et al.\ 2018, arXiv:1806.09282 

\bibitem[Wu \& Murray(2003)]{Wu2003} Wu, Y., \& Murray, N.\ 2003, \apj, 589, 605 
\bibitem[Yee et al.(2018)]{yee18} Yee, S.~W., Petigura, E.~A., Fulton, B.~J., et al.\ 2018, \aj, 155, 255 

\bibitem[Zechmeister \& K{\"u}rster(2009)]{zk09} Zechmeister, M., \& K{\"u}rster, M.\ 2009, \aap, 496, 577 

\bibitem[Zechmeister et al.(2013)]{z13} Zechmeister, M., K{\"u}rster, M., Endl, M., et al.\ 2013, \aap, 552, A78 

\end{thebibliography}




\bsp	
\label{lastpage}
\end{document}